%% file: prl_2nd.tex
\documentclass[
aps,
prl,
reprint,
superscriptaddress,
nofootinbib,
longbibliography
]{revtex4-2}

\usepackage{graphicx}
\usepackage{multirow}
\usepackage{amsmath,amssymb,amsfonts}
\usepackage{mathrsfs}
\usepackage{xcolor}
\usepackage{textcomp}
\usepackage{array}
\usepackage{booktabs}
\usepackage{comment}
\usepackage{soul}
\usepackage{hyperref}
\usepackage{bm}
\usepackage[modulo]{lineno}
\usepackage{subfigure}
\usepackage{orcidlink}

\begin{document}

\title{A First Indication of Low-Energy Solar Neutrino–Electron Scattering with PandaX-4T Run 2 Data}

\input{authorlist_20260601.tex}

\date{\today}

\begin{abstract}
We report a search for the solar proton--proton ($pp$) neutrino flux via neutrino--electron elastic scattering using the PandaX-4T Run 2 data set collected between 2024 and 2026, corresponding to an exposure of 1.9 tonne$\cdot$yr.
Before Run 2 data taking, the detector underwent a series of upgrades to improve its response and background conditions.
Time variations of radioactive noble-gas impurities are constrained using the physics data themselves, complemented by measurements from the gas-assay system.
The analysis introduced improvements in the data processing chain, detector response characterization, and background models.
A blind spectral analysis was then performed on the electronic-recoil data across a wide energy range from 20 to 1000 keV.
In combination with the Run 0 data published earlier, the pp neutrino flux is derived to be $(8.5 \pm 3.5)\times 10^{10}$ $\mathrm{cm^{-2}s^{-1}}$, consistent with the prediction of the Standard Solar Model. 
The combined dataset yields a $2.2\sigma$ statistical significance for solar neutrino-electron scattering, representing the first positive indication obtained using data below the Borexino analysis threshold of 165~keV.

\end{abstract}
\maketitle
Since Davis's pioneering experiment~\cite{Davis:1968cp}, solar neutrinos have provided a unique window into the nuclear fusion processes sustaining the Sun.
The subsequent, longstanding solar electron-neutrino deficit~\cite{Cleveland:1998nv,GALLEX:1998kcz,SAGE:1999nng,KamiokandeSolar1989,SuperKamiokandeSolar1998} was resolved through the discovery of neutrino oscillations~\cite{Super-Kamiokande:1998kpq,SNO:2002tuh}.
This discovery now stands as unambiguous evidence of nonzero neutrino mass, further demonstrating the incompleteness of the Standard Model of particle physics.
The $pp$ branch is the primary source of solar neutrinos, directly corresponding to the first step of hydrogen burning in the Sun~\cite{Vinyoles:2016djt}.
Measurements of $pp$ neutrinos therefore allow for precise tests of the Standard Solar Model (SSM) and the solar luminosity constraint.
Because the $pp$ endpoint is approximately $420~\mathrm{keV}$, the $pp$ neutrino oscillation probability is predominantly governed by vacuum mixing.
Consequently, measurements of the $pp$ flux complement higher-energy solar-neutrino observations and probe the transition region of the Mikheyev-Smirnov-Wolfenstein effect~\cite{Wolfenstein:1977ue, Mikheyev:1985zog}. The lower-energy solar neutrinos are not subject to it, while the higher-energy neutrinos are, with a transition occurring at a few MeV.
Furthermore, they provide a critical test of potential physics beyond the Standard Model, such as anomalous electromagnetic properties~\cite{Giunti:2014ixa}, nonstandard interactions~\cite{Ohlsson:2012kf, Farzan:2017xzy}, sterile-neutrino admixtures~\cite{Abazajian:2012ys, Palazzo:2013me}, or other scenarios that could alter the low-energy spectrum.

Direct $pp$ neutrino measurements are experimentally demanding because the resulting electronic-recoil spectrum is restricted to low energies below a few hundred keV, with a smooth, continuous shape lacking sharp spectral features.
This makes it challenging to distinguish the signal from intrinsic radioactive backgrounds within this energy range.
Liquid scintillator detectors, most notably Borexino~\cite{BOREXINO:2018ohr}, have achieved high-significance measurements
with the lowest analysis threshold of 165 keV
by leveraging large fiducial masses and exceptional radiopurity.
As we demonstrate here, liquid-xenon (LXe) time projection chambers (TPCs)~\cite{Aprile:2009dv} provide a complementary approach.
With their low energy thresholds, sizable homogeneous targets, three-dimensional position reconstruction, and strong self-shielding and background control capabilities, LXe TPCs offer distinct advantages for detecting low-energy solar neutrinos.

In this Letter, we present an analysis of solar $pp$ neutrinos using the Run 2 data from the PandaX-4T LXe detector, with a total exposure of $1.9~\text{tonne}\cdot\text{yr}$, representing a threefold increase in exposure compared to the published Run 0 results~\cite{PandaX:2024jjs}.
Run 2 benefits from several upgrades to the detector, electronics, and operational conditions of the distillation system for noble-impurity (Kr/Ar/Rn) removal.
Improvements in the
statistics,
data selection strategy, detector response models, and background models significantly enhance the sensitivity to low-energy solar neutrinos. A staged unblinding procedure is also adopted to reduce the risk of bias during analysis development.

PandaX-4T is an LXe TPC located at the second phase of China Jinping Underground Laboratory (CJPL-II)\cite{Cheng:2017usi}.
Particle interactions within the active xenon volume produce a prompt scintillation signal ($S1$) and an ionization signal.
The ionization electrons are drifted upward, extracted into the gas phase, and converted into a proportional scintillation signal ($S2$).
The drift time, combined with the hit pattern of the $S2$ signal, enables three-dimensional event reconstruction, which allows for fiducialization and the rejection of surface and external backgrounds.
Although PandaX-4T was originally designed to search for dark matter, its low electronic-recoil background and multi-ton xenon target are also well-suited for studies of solar neutrinos~\cite{PandaX:2024jjs}, double-beta decay~\cite{PandaX:2023ggs, PandaX:2025yly}, and other rare processes.

Run 2 refers to the PandaX-4T data collected from July 5, 2024, to January 23, 2026, corresponding to 438.0 live days after data-quality selections.
Prior to Run 2, the detector underwent a series of upgrades during the construction of CJPL-II.
These upgrades included the replacement of the cathode, redesigned voltage-dividing bases of the top and bottom photomultiplier tubes (PMTs) to improve their linear response, and new polytetrafluoroethylene (PTFE) field-cage panels with enhanced photon reflectivity and reduced tritium attachment.
In addition, the external ultrapure-water shield was upgraded into an active water Cherenkov veto system instrumented with 270 8-inch PMTs.
Although this system provides efficient tagging of cosmic-ray muons and associated backgrounds, its veto information is not used in the present analysis.

The readout electronics have also been upgraded.
All TPC channels are now read out using custom-designed 14-bit digitizers operating at a sampling rate of 500 MS/s~\cite{He:2021digitizer}, doubling the previous rate.
The channel-level threshold is approximately one-third of a photoelectron.
During physics data taking, the digitizers operate in a triggerless mode, wherein over-threshold waveforms are read out with baseline suppression and subsequently clustered by a dedicated event-building computer.
During calibration campaigns, in which the data rate is significantly higher, the field-programmable gate array (FPGA) performs real-time digital summation of individual channels and generates external triggers, either prescaled or restricted to a predefined energy window, thereby reducing the data throughput and minimizing event pileup.

The liquid xenon is continuously circulated through a purification system consisting of hot zirconium getters to remove electronegative impurities.
Due to the two parallel recirculation loops, the recirculation flow rate is varied between 40 and 60 standard liters per minute, where one standard liter is the amount of gas occupying 1\, L at atmospheric pressure and 0\,°C.
Throughout Run 2, an average electron lifetime of $1.08 \pm 0.05\,\mathrm{ms}$ was achieved, ensuring efficient charge collection over the full TPC drift length.

The actual ${}^{222}$Rn activity varies due to changing detector operational conditions.
To suppress the radon concentration in the liquid xenon, xenon extracted from the recirculation loops was routed through the distillation tower before entering the hot getter.
Unlike its standard operation for krypton and argon removal, the distillation system was operated in a dedicated ``reverse mode'' optimized for radon reduction~\cite{Cui:2024ltd}.
As a result, the ${}^{222}$Rn activity in the TPC was reduced to a minimum of $2.9 \pm 0.1\,\mu\mathrm{Bq/kg}$, substantially lower than the average activity of $8.7 \pm 0.3\,\mu\mathrm{Bq/kg}$ reported in the previous PandaX-4T analysis~\cite{PandaX:2024qfu}.
The reduced radon activity suppresses one of the dominant low-energy electronic-recoil backgrounds, thereby improving the sensitivity to solar $pp$ neutrinos.

The $^{39}$Ar concentration was found to be significantly elevated compared to previous data sets~\cite{PandaX:2024cic}.
To address this, two online argon-distillation campaigns, lasting approximately 39 and 34 days, respectively, were conducted by operating the distillation tower in normal mode during data taking.
The Run 2 data are therefore separated into a \emph{High-Ar} period with 233.8 live days and a \emph{Low-Ar} period of 204.2 live days.
The corresponding average $^{39}$Ar activities are $0.74~\mu\mathrm{Bq/kg}$ and $0.10~\mu\mathrm{Bq/kg}$, respectively, as described in the background model below.

\begin{figure}
\centering
\includegraphics[width=1.\linewidth]{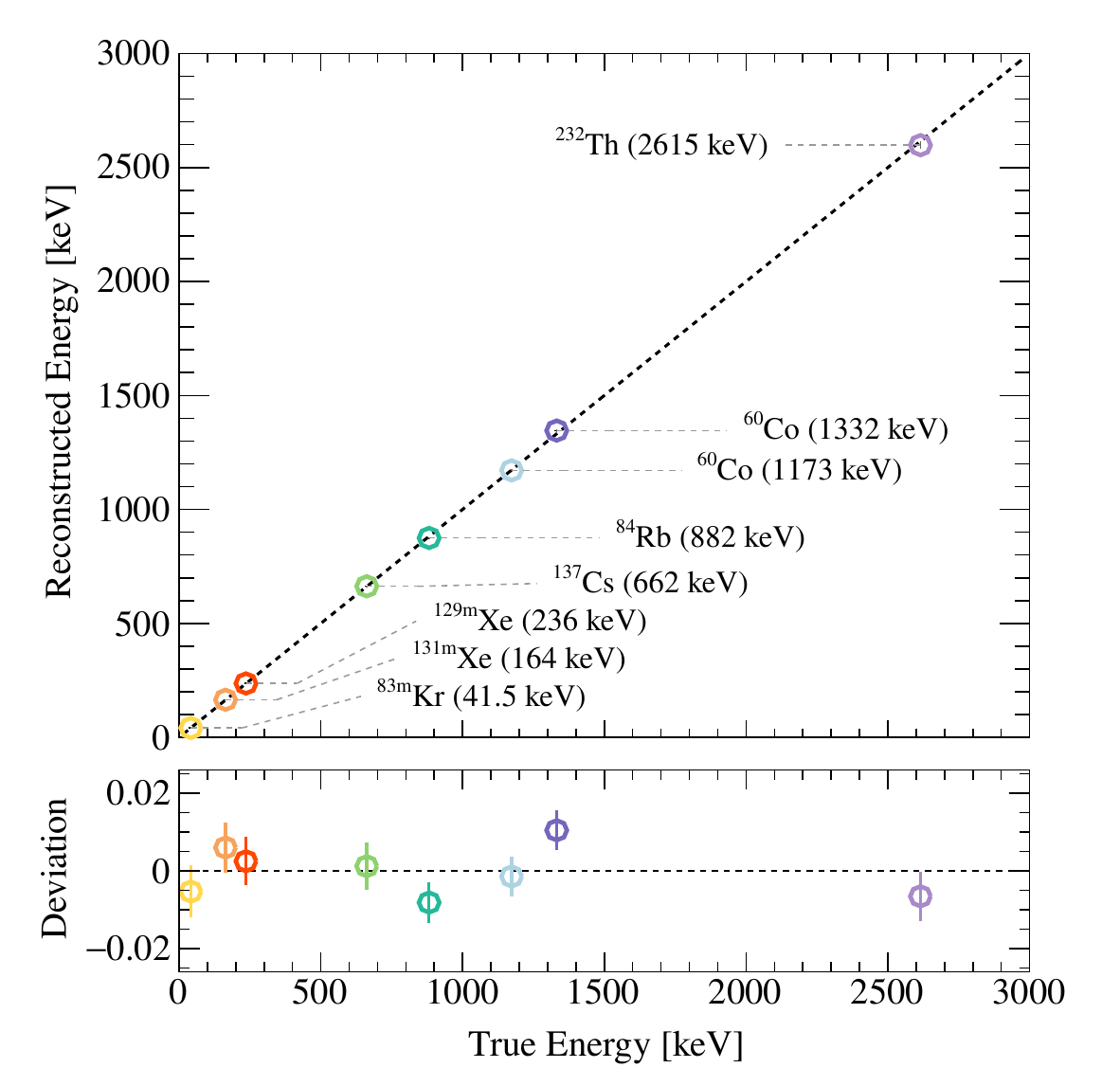}
\caption{Reconstructed energy (top) and relative deviation (bottom) vs. true energy for monoenergetic peaks from calibration sources in PandaX-4T Run 2. The deviation is defined as (reconstructed energy $-$ true energy) / true energy.}
\label{fig:g1g2b}
\end{figure}

The data processing chain was substantially revised for Run 2.
In previous PandaX-4T analyses, low-energy dark matter searches (below approximately 25~keV) and the neutrino analysis (spanning hundreds of keV to 2.8~MeV) were processed using separate reconstruction pipelines optimized for their respective energy ranges.
For this analysis, the pipelines were merged into a single reconstruction chain with common algorithms and configuration parameters throughout all major processing stages.
At very low energies, the new chain can now identify $S1$ signals down to a single PMT hit, reducing the energy threshold.
At the higher energies relevant to this analysis, we implemented improvements to better identify the pileup of multiple $S2$ signals, enabling a more accurate computation of the associated charge and position.
The unified chain eliminates systematic uncertainties that could arise from differences between the two separate processing pipelines.

Event selection is performed after identifying the $S1$ and $S2$ signals.
The event-selection efficiency is determined using a $^{232}$Th calibration source placed outside the cryostat at the middle of the TPC.
The measured spectrum is compared with a Geant4-based simulation, which utilizes the same detector geometry and radioactive decay model as in previous PandaX-4T analyses.
The simulation records energy depositions in the active target and applies a simplified detector-response model, the single-site selection, and energy-resolution smearing.
The ratio between the measured and simulated spectra yields the selection efficiency as a function of reconstructed energy.
An overall uncertainty of 6.7\% is assigned to this efficiency curve.

For Run 2, the upgraded PMT bases extended the upper limit of the PMT linearity range from 1000 photoelectrons (PE) up to 40000 PE~\cite{Luo:2023ebw}.
This enhancement significantly improves the energy reconstruction performance.
The observable energy is reconstructed from the corrected $S1$ and bottom $S2$ signals, denoted as $Q_{S1}^{\rm c}$ and $Q_{S2_{\rm b}}^{\rm c}$, respectively.
The superscript ``c'' indicates corrections based on spatial-uniformity maps derived from the monthly $^{\rm 83m}$Kr calibration data.
The reconstructed energy is given by
\begin{equation}
\label{eq:doke}
E = W\left(\frac{Q_{S1}^{\rm c}}{g_1}+\frac{Q_{S2_{\rm b}}^{\rm c}}{g_{2_{\rm b}}}\right),
\end{equation}
where $W=13.7$~eV is the average work function in liquid xenon, $g_1$ is the photon-detection gain, and $g_{2_{\rm b}}$ is the charge-amplification gain.
For multi-site events, the summed and corrected $S2_{\rm b}$ signal is used for energy reconstruction.
The values of $g_1=0.13$ and $g_{2_{\rm b}}=8.61$ are determined via fitting the full absorption peaks of 41.5 keV ($^{\mathrm{83m}}$Kr), 164 keV ($^{\mathrm{131m}}$Xe), 236 keV ($^{\mathrm{129m}}$Xe), 662 keV ($^{\mathrm{137}}$Cs), 882 keV ($^{\mathrm{84}}$Rb), 1173 keV ($^{\mathrm{60}}$Co), 1332 keV ($^{\mathrm{60}}$Co), and 2615 keV ($^{\rm 232}$Th) using Eq.~\ref{eq:doke}.
Excellent energy linearity is achieved throughout the entire energy range up to 2.6 MeV, as demonstrated in Fig.~\ref{fig:g1g2b}.

To fit the final physics data, we construct a five-parameter detector energy response model
following the previous analysis~\cite{PandaX:2024sds}
: the fractional energy resolution is parameterized as $\frac{\sigma(E)}{E} = \frac{a}{\sqrt{E}}+b \cdot E + c$, where $E$ is the reconstructed energy in keV; and the residual energy nonlinearity is modeled as $E=d \cdot \hat{E} + e$, where $\hat{E}$ is the true energy.

The prior values, uncertainties, and covariance matrix of these parameters are obtained from a fit to monoenergetic peaks in the physics data within a peripheral zone outside the 1.80-tonne central volume.

The dominant backgrounds in the signal region are electronic recoils from three main sources: radioactivity in detector materials, noble-gas impurities dissolved in the xenon target, and radioactive xenon isotopes.
Detailed breakdowns can be seen in Table~\ref{tab:bkg}.

\begin{table}[tbp]
\caption{Background contributions expected in the ROI and the fitted counts obtained from the log-likelihood fit. Here HAr and LAr denote the High-Ar and Low-Ar data sets, respectively.}
\label{tab:bkg}
\centering
\begin{tabular}{lcc}
\toprule
Components & \quad\quad Expected ($\times 10^2$) \quad\quad & Fitted ($\times 10^2$) \\
\midrule
DetM $^{60}$Co & 85 $\pm$ 25 & 115 $\pm$ 18 \\
DetM $^{40}$K & 134 $\pm$ 28 & 149 $\pm$ 22 \\
DetM $^{232}$Th & 656 $\pm$ 33 & 571 $\pm$ 27 \\
DetM $^{238}$U & 159 $\pm$ 64 & 186 $\pm$ 15 \\
SSP $^{232}$Th & 234 $\pm$ 21 & 215 $\pm$ 22 \\
SSP $^{238}$U & (2.7 $\pm$ 1.3)$\times10^2$ & 188 $\pm$ 66 \\
$^{85}$Kr (HAr) & 63 $\pm$ 14 & 72 $\pm$ 13 \\
$^{85}$Kr (LAr) & 41 $\pm$ 12 & 35.9 $\pm$ 7.1 \\
$^{39}$Ar (HAr) & (2.3 $\pm$ 1.2)$\times10^2$ & 139 $\pm$ 18 \\
$^{39}$Ar (LAr) & 27 $\pm$ 11 & 41.9 $\pm$ 7.6 \\
$^{214}$Pb (HAr) & free & 492 $\pm$ 23 \\
$^{214}$Pb (LAr) & free & 510 $\pm$ 23 \\
$^{212}$Pb & 24 $\pm$ 11 & 49.8 $\pm$ 6.0 \\
$^{136}$Xe & 1499 $\pm$ 37 & 1568 $\pm$ 57 \\
$^{124}$Xe & 5.71 $\pm$ 0.87 & 6.94 $\pm$ 0.55 \\
$^{133}$Xe (HAr) & 49.8 $\pm$ 7.5 & 55.3 $\pm$ 3.7 \\
$^{133}$Xe (LAr) & (7.5 $\pm$ 5.2)$\times10^{-2}$ & (7.7 $\pm$ 5.4)$\times10^{-2}$ \\
Xe$^{*}$ (HAr) & free & 1025 $\pm$ 29 \\
Xe$^{*}$ (LAr) & free & 64.0 $\pm$ 2.1 \\
solar $pp$ neutrino & free & 12.9 $\pm$ 5.7 \\
\bottomrule
\end{tabular}
\end{table}

Radioactivity from detector materials is categorized into two classes.
The first class, denoted DetM, includes detector components inside or near the TPC.
The second class describes the stainless-steel platform (SSP) that supports the xenon cryostat within the water shield.
For DetM, we simulate $^{238}$U, $^{232}$Th, ${}^{40}$K, and ${}^{60}$Co decays uniformly within each material, setting the initial normalizations to the central values from high-purity germanium assay measurements~\cite{PandaX-4T:2021lbm}.
We then sum the material contributions by isotope group, leaving four floating normalization parameters.
For the SSP component, which was first identified in a previous multi-site analysis~\cite{PandaX-4T:2025jel}, we generate $^{238}$U and $^{232}$Th decays uniformly in the SSP, incorporating two additional normalization parameters.
These six parameters are constrained by fitting the multi-site data set, where gamma-ray backgrounds are significantly enhanced.
This process is repeated for various fiducial volume selections, enabling the determination of the six parameters and their systematic uncertainties, which are then propagated into the single-site fit.

The $^{39}$Ar background was identified via the enhanced Ar partial pressure in xenon using a residual gas analyzer system~\cite{Leonard:2001rpt}.
Based on the evolution of the Ar level during the distillation campaign, the $^{39}$Ar activity obtained via energy spectral fits scales linearly with the partial pressure measurement, as expected.
The $^{39}$Ar levels in the high-Ar and low-Ar data sets are estimated to be $0.74 \pm 0.39 ~\mu\mathrm{Bq/kg}$ and $0.10 \pm 0.04 ~\mu\mathrm{Bq/kg}$, respectively, with uncertainties uncorrelated between the two data sets.

The $^{85}$Kr background is estimated using the same delayed-coincidence strategy as in earlier PandaX-4T analyses~\cite{PandaX:2024qfu}.
In the High-Ar and Low-Ar data sets, 20 and 13 $^{85}$Kr$\rightarrow ^{\mathrm{85m}}$Rb delayed-coincidence candidates are identified, corresponding to molar krypton concentrations in xenon of $1.10\pm0.25$~part per trillion (ppt) mol/mol and $0.82\pm0.23$~ppt mol/mol, respectively.

The $^{222}$Rn and $^{220}$Rn chains are identified through alpha decays in the detector, with $^{214}$Pb and $^{212}$Pb constituting the dominant beta backgrounds.
Due to its distinct spectral feature, the $^{214}$Pb component is treated as a free parameter in both data sets.
The activity of $^{220}$Rn remains relatively stable, and a shared $^{212}$Pb activity of 0.051 $\pm$ 0.024 $\mathrm{\mu Bq/kg}$ is estimated based on the radioactivity of $^{212}$Po and $^{220}$Rn.

The xenon isotopic background includes contributions from $^{136}$Xe, $^{124}$Xe, $^{125}$Xe, $^{127}$Xe, $^{129\mathrm{m}}$Xe, $^{131\mathrm{m}}$Xe, and $^{133}$Xe.
The contributions from $^{136}$Xe two-neutrino double-beta decay and $^{124}$Xe double-electron capture are calculated using the half-lives measured previously by PandaX-4T~\cite{PandaX:2025yly, PandaX-4T:2024fls}.
The amplitudes of monoenergetic peaks from $^{125}$Xe, $^{127}$Xe, $^{129\mathrm{m}}$Xe, and $^{131\mathrm{m}}$Xe, such as those at 164, 236, 380, and 408 keV, are left unconstrained in the fit.
$^{133}$Xe ($T_{1/2} = 5.25\text{ d}$) features
an 81 keV de-excitation $\gamma$-ray in combination with a continuous $\beta$-energy.
The $^{133}$Xe activities in the High-Ar and Low-Ar data sets are 0.15 $\pm$ 0.02 $\mathrm{\mu Bq /kg}$ and (2.7 $\pm$ 1.9) $\times$ 10${^{-4}}$ $\mu\mathrm{Bq /kg}$, respectively, estimated based on the time evolution since neutron calibration in the single-site data.

The solar-neutrino signal model includes $pp$ and ${}^{7}$Be neutrino-electron scattering spectra calculated from the standard solar flux spectra, oscillation probabilities, and the electroweak scattering cross section~\cite{Chen:2016eab}.
Following Ref.~\cite{Chen:2016eab}, the low-energy neutrino-ionization cross section incorporates xenon atomic binding and relativistic many-body effects through the relativistic random-phase approximation, which suppresses the recoil rate relative to the free-electron approximation.
These spectra are convolved with the detector energy response and multiplied by the selection efficiency.
We introduce a normalization parameter, $\mu$, representing the overall scale factor of the combined solar $pp$+$^{7}\text{Be}$ neutrino spectrum relative to the SSM prediction. The fitted value of $\mu$ is
used to derive a $pp$-only flux under the SSM-predicted relative normalization of the $pp$ component.

\begin{figure}[tbp]
\centering
\subfigure{\label{fig:xydis}\includegraphics[width=0.4\textwidth]{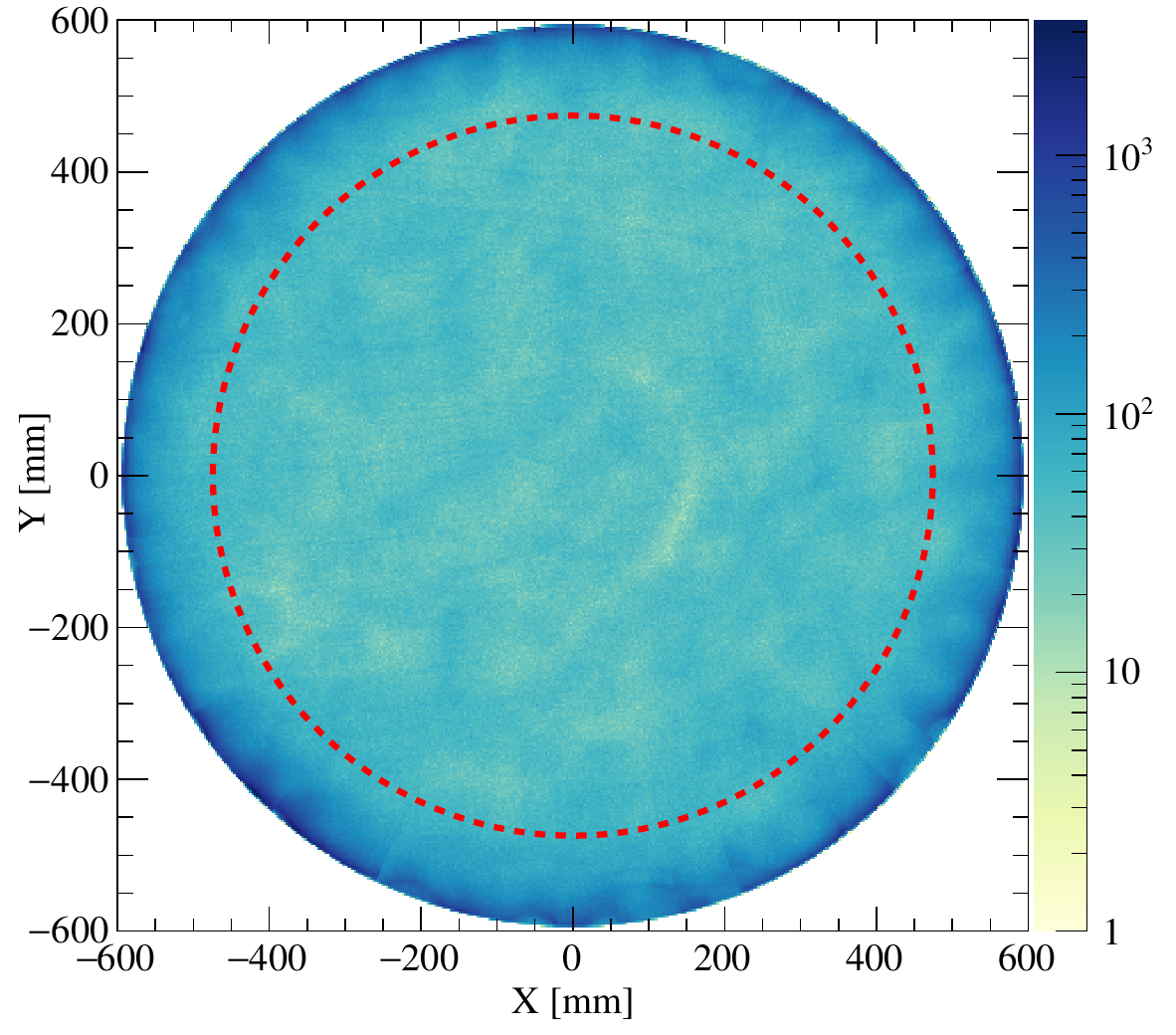}}
\\[ -0.2cm ]
\subfigure{\label{fig:zr2dis}\includegraphics[width=0.4\textwidth]{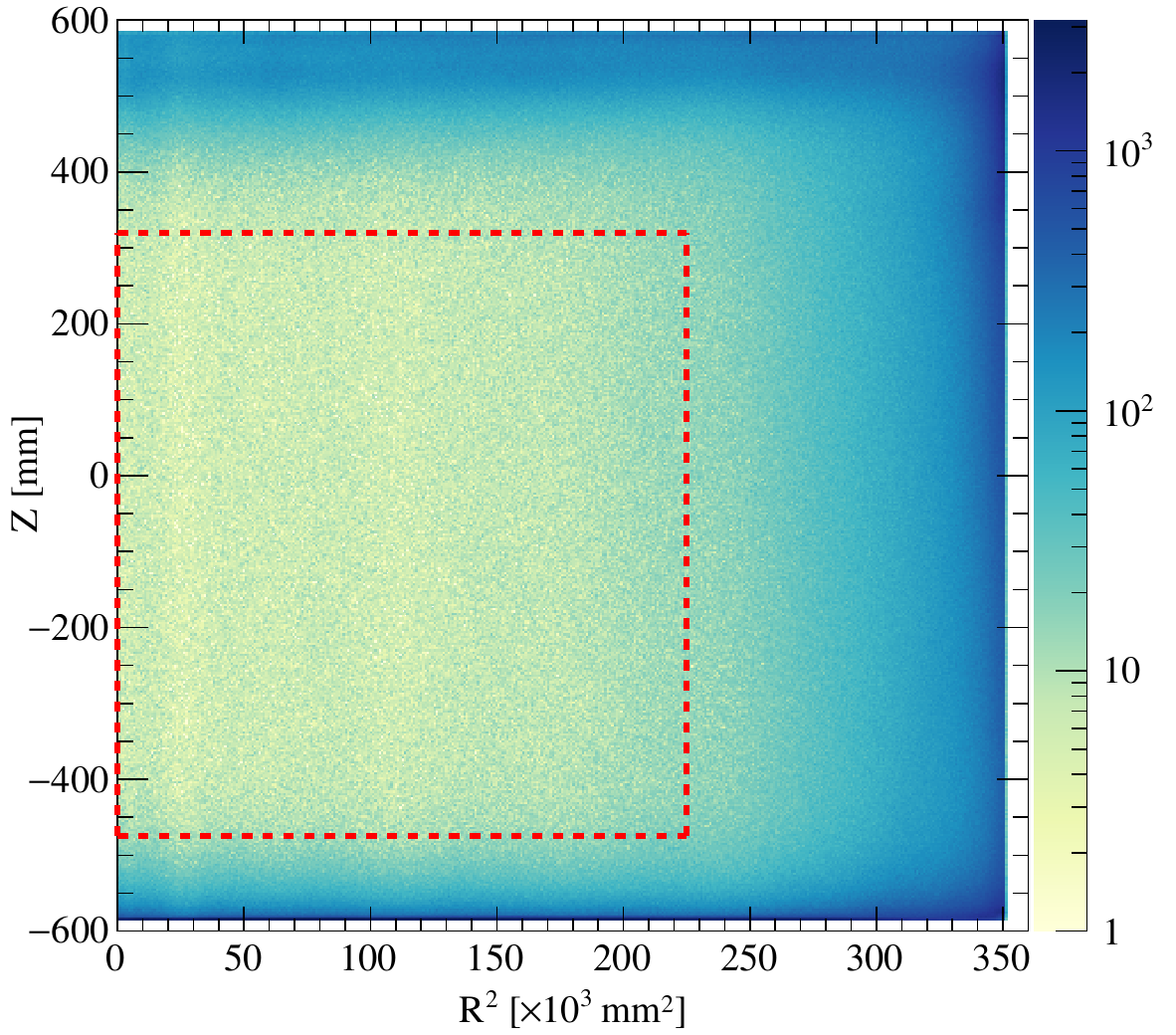}}
\caption{Spatial distributions of the selected physics events in the \(X\)–\(Y\) (top) and \(R^2\)–\(Z\) (bottom) projections. The color scale shows the event density on a logarithmic scale. The red dashed circle (top) and box (bottom) mark the fiducial-volume boundaries, corresponding to a fiducial mass of \(1.60\pm0.03\) tonnes.
}
\label{fig:distribution}
\end{figure}

Once the measured energy spectrum and the signal and background models are established, the data are analyzed via
a binned-likelihood fit following a staged unblinding procedure.

The High-Ar and Low-Ar data sets are fitted simultaneously.
The negative log-likelihood is constructed following Ref.~\cite{PandaX:2024sds}, consisting of Poisson terms for the observed counts in each energy bin and constraint terms for nuisance parameters.
The latter terms account for uncertainties in the overall efficiency and externally constrained background models.
The signal normalization parameter $\mu$ is left floating, common to both the High-Ar and Low-Ar data sets.
Except for $^{220}\text{Rn}$, $^{136}\text{Xe}$, $^{124}\text{Xe}$, and radioactivities of DetM and SSP, all other background components are treated independently in both data sets.
For each signal or background component, the energy spectrum is convolved with the five-parameter energy response model described above.
The same detector-response parameters are applied to both data sets in the joint fit.

The fit range and fiducial-volume selection were finalized through a multi-stage blind analysis.
The lower boundary of 20 keV is chosen to preserve the blinding of the lower-energy region for the dark matter search.
An initial 10\% of the data, randomly sampled over the full run period, was unblinded to validate the event selections, background models, data simulation agreement, and trial fitting procedure.
Based on the projected fractional uncertainty of the $pp$ signal, a fit range of 20--2000 keV and a fiducial mass of 1.80 tonne were initially selected according to the expected sensitivity to the
common $pp$+$^{7}\mathrm{Be}$ normalization.
A second 10\% data sample was then opened, with no significant anomaly observed.
After unblinding an additional 30\% of the data,
we performed a cross-check with the background spectra used in the published Run 0 analysis~\cite{PandaX:2024sds}. For the simulated energy spectra of backgrounds from noble gas impurities and activated xenon isotopes in this analysis, we realized that the fiducial-volume cut applied to the experimental data had been inadvertently omitted from the simulated data. After fixing this inconsistency in the simulated background spectra, we re-optimized the fit range and fiducial volume with a figure of merit combining the expected pp-signal uncertainty and the goodness-of-fit under the background+SSM hypothesis.
This procedure led to a final region of interest of 20--1000 keV and a fiducial mass of $1.60 \pm 0.03$ tonnes.
The remaining 50\% of the data were then unblinded, yielding approximately $5.5\times10^{5}$ selected events, as shown in Fig.~\ref{fig:distribution}.

\begin{figure}[tbp]
\centering
\subfigure{\label{fig:higharfit}\includegraphics[width=1.\linewidth]{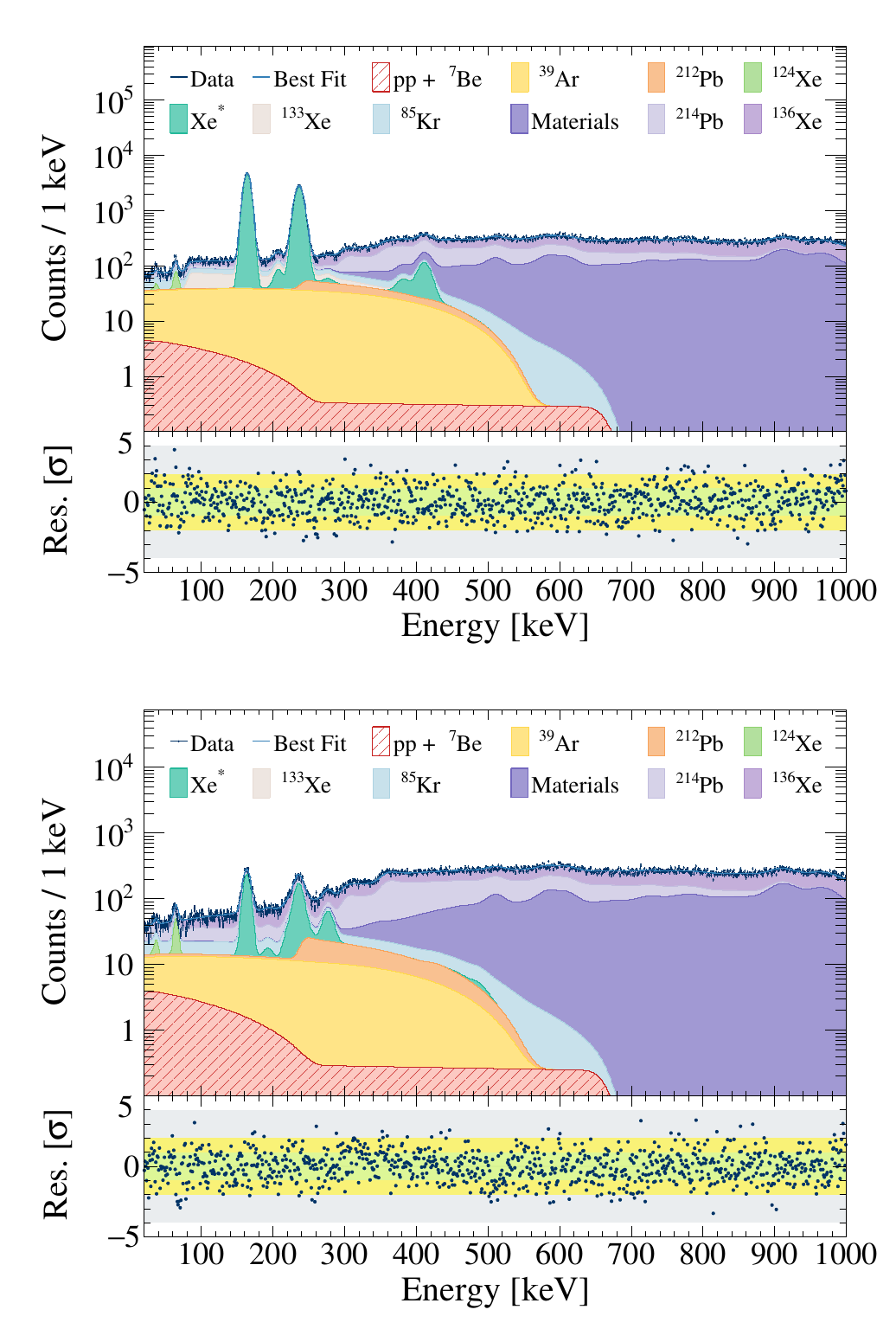}}
\caption{Spectral fits to the High-Ar (top) and Low-Ar (bottom) data sets in the 20–1000 keV energy range after unblinding. The bin width is 1 keV. Xe$^{*}$ denotes the sum of all xenon-isotope contributions except ${}^{124}$Xe, ${}^{133}$Xe and ${}^{136}$Xe. The lower panels show the fit residuals alongside the $\pm1\sigma$, $\pm2\sigma$, and $\pm4\sigma$ bands.}
\label{fig:fitresult}
\end{figure}

\begin{table}[tbp]
\centering
\caption{Systematic uncertainty contributions to the solar $pp+{}^{7}\mathrm{Be}$ neutrino normalization factor $\mu$. "Overall efficiency" accounts for uncertainties in the quality-cut acceptance and single-site selection common to signal and backgrounds, while “number of target electrons” accounts for uncertainties in the fiducial-volume definition and liquid-xenon density affecting the signal normalization. For each constrained nuisance parameter, the contribution is estimated from the change in the uncertainty of $\mu$ when the corresponding penalty term is removed from the likelihood. It should be noted that individual contributions are highly correlated in the fit, and the quadrature sum does not reproduce the total systematic uncertainty. The contribution associated with $^{214}\mathrm{Pb}$, evaluated in this manner, is 0.21. However, since $^{214}\mathrm{Pb}$ is not constrained in the fit, its contribution is already accounted for in the statistical uncertainty and therefore not presented in the table below.}

\label{tab:err}
\begin{tabular}{l l c}
\toprule
\multicolumn{2}{l}
{Component} & Contribution
\\\midrule\multicolumn{2}{l}
{Overall efficiency} & 0.26
\\\multicolumn{2}{l}{
Number of target electrons} & 0.04
\\\multicolumn{2}{l}{Detector response} & 0.13
\\\midrule
\multirow{6}{*}{Background \quad\quad}
& $^{39}$Ar & 0.39 \\
& $^{85}$Kr & 0.38 \\
& Material\quad\quad & 0.36 \\
& $^{136}$Xe & 0.20 \\
& $^{124}$Xe & 0.15 \\
& $^{212}$Pb & 0.14 \\
& $^{133}$Xe & 0.09 \\
\bottomrule
\end{tabular}
\end{table}

The measured energy spectra for the High-Ar and Low-Ar data sets are shown alongside the best-fit results in Fig.~\ref{fig:fitresult}, yielding a $\chi^2$/ndf = 2312.3/1946 = 1.19.
The post-fit nuisance parameters remain consistent with their pre-fit constraints within their assigned uncertainties.

The fit yields a solar $pp$+$^{7}\mathrm{Be}$ neutrino normalization of $\mu = 1.42 \pm 0.63$, consistent with the SSM prediction.
By fixing all nuisance parameters to their best-fit values, the statistical uncertainty is isolated to be 0.22.
The systematic uncertainty is then obtained by subtracting the statistical component from the total uncertainty in quadrature, giving 0.59.
The individual systematic contributions are summarized in Table~\ref{tab:err}.
The dominant uncertainties arise from ${}^{85}\mathrm{Kr}$ and ${}^{39}\mathrm{Ar}$, reflecting their strong spectral degeneracy with the solar $pp$ neutrino signal.
In addition, the uncertainty in the predicted \(^{7}\mathrm{Be}\) flux~\cite{Vinyoles:2016djt}, the alternative spectral shapes for \(^{39}\mathrm{Ar}\), \(^{85}\mathrm{Kr}\), and \(^{212}\mathrm{Pb}\)~\cite{Nitescu:2023sry,Haselschwardt:2020iey}, and the use of separate energy-response models for the High-Ar and Low-Ar data each have a negligible impact on the extracted $pp$ neutrino flux relative to the total measurement uncertainty.

\begin{figure}[htbp]
\centering
\includegraphics[width=1.\linewidth]{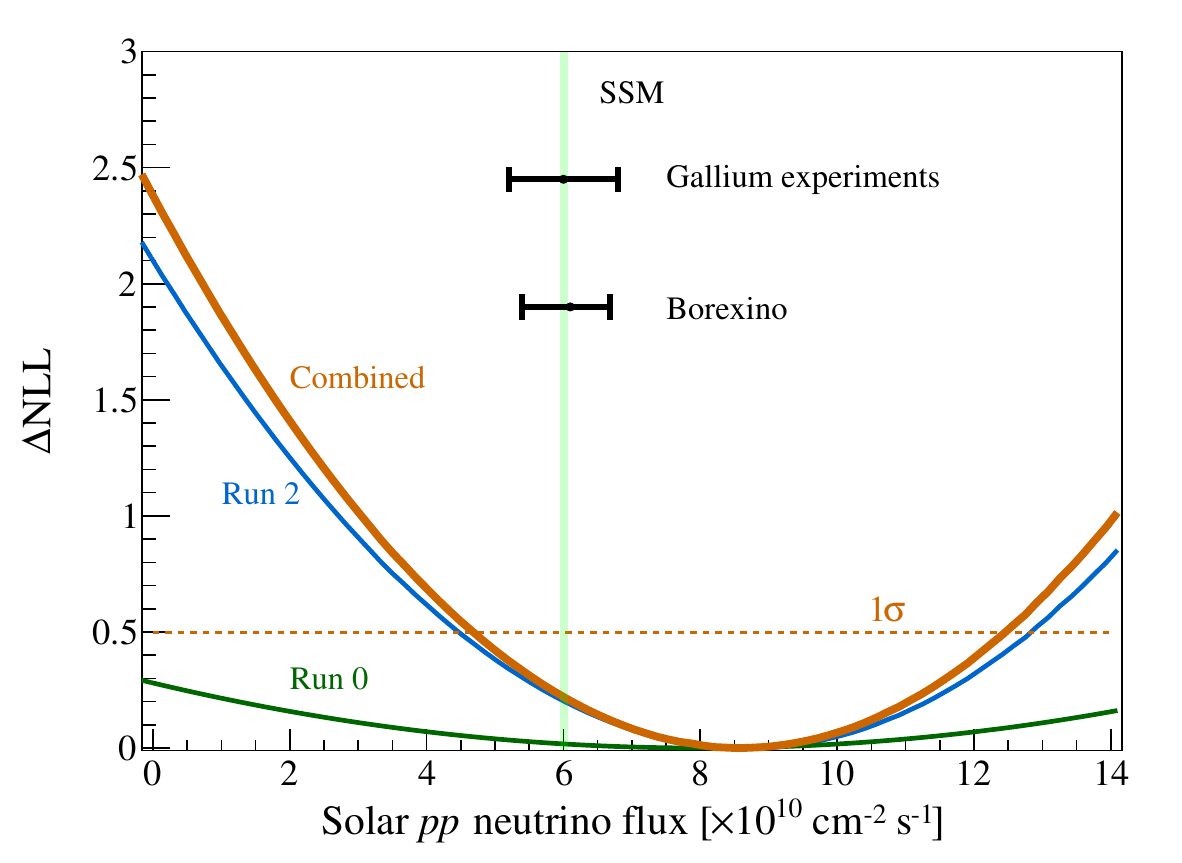}
\caption{Observed $\Delta$NLL (negative log likelihood) vs. solar $pp$ neutrino flux. The inferred pp-only fluxes from Run 2 and the combined Run 0+2 results are $(8.5 \pm 3.7)\times10^{10}\,\mathrm{cm}^{\mathrm{-2}}\mathrm{\cdot s}^{\mathrm{-1}}$ and $\mathrm {(8.5 \pm 3.5)}\times10^{10}$ $\mathrm{cm}^{\mathrm{-2}}\mathrm{\cdot s}^{\mathrm{-1}}$, respectively. The shaded green band indicates the SSM predictions based on the high-metallicity (GS98)~\cite{Grevesse:1998bj} and low-metallicity (AGSS09)~\cite{Asplund:2009fu} models. Measurements from gallium experiments~\cite{SAGE:2009eeu} and Borexino~\cite{BOREXINO:2018ohr} are shown for comparison with arbitrary vertical offsets.}
\label{fig:NLL}
\end{figure}

For comparison with other $pp$-neutrino measurements, we quote the result as an equivalent $pp$-only rate by multiplying the fitted $\mu$ with the SSM-predicted $pp$ flux $\Phi_{\rm SSM}^{pp}$ (Fig.~\ref{fig:fitresult}).
The corresponding $pp$-only flux inferred under the SSM assumption
is $(8.5 \pm 3.7)\times10^{10}\,\mathrm{cm}^{\mathrm{-2}}\mathrm{\cdot s}^{\mathrm{-1}}$, where the uncertainty includes both statistical and systematic components.
Combining this result with the published Run 0
result
~\cite{PandaX:2024jjs} reduces the uncertainty slightly to $(8.5 \pm 3.5)\times10^{10}\,\mathrm{cm}^{\mathrm{-2}}\mathrm{\cdot s}^{\mathrm{-1}}$, also consistent with the SSM prediction~\cite{Vinyoles:2016djt} within 1$\sigma$.
To quantify the statistical significance, we perform a profile-likelihood scan (Fig.~\ref{fig:NLL}). Combined with Run 0, the data disfavor the zero-$pp$ hypothesis at $2.2\sigma$.

Utilizing only a modest 1.9~tonne$\cdot$yr exposure, this
result
establishes liquid xenon detectors as a complementary platform for low-energy solar neutrino physics, yielding results with systematic uncertainties entirely distinct from those of gallium and liquid scintillator experiments.

In summary, we have searched for low-energy solar neutrino–electron scattering using PandaX-4T data.
Under the SSM normalization of the $pp$ + ${}^{7}$Be components, the result corresponds to a $pp$ flux of $(8.5 \pm 3.5)\times 10^{10}~\mathrm{cm^{-2}s^{-1}}$, consistent with the SSM prediction.
With a statistical significance of $2.2\sigma$, this marks the first positive indication of low-energy solar neutrino–electron scattering using the data below Borexino’s analysis threshold of 165~keV~\cite{BOREXINO:2018ohr}.
Present sensitivity is limited by statistics and, more critically, by spectral degeneracies between the smooth $pp$ recoil spectrum and low-energy radioactive backgrounds.
Overcoming these limitations requires both larger exposures and targeted reductions of $^{39}\text{Ar}$, $^{85}\text{Kr}$, and radon backgrounds, alongside an external veto detector to mitigate material-induced backgrounds.
With these intrinsic backgrounds under control, future multi-tonne liquid xenon TPCs, such as PandaX-xT~\cite{PANDA-X:2024dlo}, will further extend this multi-purpose capability, simultaneously serving as precision probes of low-energy solar neutrinos and pushing sensitivities to dark matter and other rare processes to new frontiers.

We thank Ovidiu Ni{\c{t}}escu for providing the ${}^{39}\mathrm{Ar}$ and ${}^{85}\mathrm{Kr}$ $\beta$-decay spectra calculated following the method of Ref.~\cite{Nitescu:2023sry}, and for valuable discussions.
This project is supported in part by grants from the National Key R\&D Program of China (No. 2023YFA1606200, No. 2023YFA1606201, and No. 2023YFA1606202), National Science Foundation of China (No. 12090060, No. 12090061, No. 12090062, No. U23B2070, No. W2543017), and by the Office of Science and Technology, Shanghai Municipal Government (Grants No. 21TQ1400218, No. 22JC1410100, No. 25ZR1402223), and by the Natural Science Foundation of Sichuan, China (Grant No. 2026NSFSC0757).
We are grateful for the support of the Fundamental Research Funds for the Central Universities.
We also thank the following for their sponsorship: the Chinese Academy of Sciences Center for Excellence in Particle Physics (CCEPP), Thomas and Linda Lau Family Foundation, New Cornerstone Science Foundation, Tencent Foundation in China, and Yangyang Development Fund.
Finally, we thank the CJPL administration and the Yalong River Hydropower Development Company Ltd. for indispensable logistical support and other help.

This work is a joint effort by the PandaX Collaboration, which constructed, operated, and processed data for the PandaX-4T apparatus.
Besides the corresponding authors who led this analysis and prepared the manuscript, key individual contributions to this work are as follows.
X. Cui was responsible for the Kr/Ar/Rn distillation operations.
Z. Gao performed the fiducial-volume optimization, spectral fitting, systematic uncertainty studies, and solar ($pp$)-signal extraction.
G. Wang produced the background and data spectra and independently validated the fitting results.
X. Chen coordinated the Run 2 data processing and the integration of the low- and high-energy analysis chains.
J. Wang evaluated the event selection and efficiency.
Y. Yun determined the detector exposure and position reconstruction.
C. Han developed the energy reconstruction and provided essential inputs for the detector-response model.

\bibliography{references}
\end{document}

%% file: authorlist_20260601.tex

\def\tdli{State Key Laboratory of Dark Matter Physics, Key Laboratory for Particle Astrophysics and Cosmology (MoE), Shanghai Key Laboratory for Particle Physics and Cosmology, New Cornerstone Science Laboratory, Tsung-Dao Lee Institute, Shanghai Jiao Tong University, Shanghai 201210, China}
\def\sjtuphys{State Key Laboratory of Dark Matter Physics, Key Laboratory for Particle Astrophysics and Cosmology (MoE), Shanghai Key Laboratory for Particle Physics and Cosmology, School of Physics and Astronomy, Shanghai Jiao Tong University, Shanghai 200240, China}
\def\MESJTU{School of Mechanical Engineering, Shanghai Jiao Tong University, Shanghai 200240, China}
\def\SPEIT{SJTU Paris Elite Institute of Technology, Shanghai Jiao Tong University, Shanghai 200240, China}
\def\SJTUSC{Shanghai Jiao Tong University Sichuan Research Institute, Chengdu 610213, China}

\def\BUAA{School of Physics, Beihang University, Beijing 102206, China}
\def\BUAACenter{Peng Huanwu Collaborative Center for Research and Education, Beihang University, Beijing 100191, China}
\def\BUAALab{International Research Center for Nuclei and Particles in the Cosmos \& Beijing Key Laboratory of Advanced Nuclear Materials and Physics, Beihang University, Beijing 100191, China}
\def\SCNT{Southern Center for Nuclear-Science Theory (SCNT), Institute of Modern Physics, Chinese Academy of Sciences, Huizhou 516000, China}

\def\USTClab{State Key Laboratory of Particle Detection and Electronics, University of Science and Technology of China, Hefei 230026, China}
\def\USTCdep{Department of Modern Physics, University of Science and Technology of China, Hefei 230026, China}

\def\YaLongSD{Yalong River Hydropower Development Company, Ltd., 288 Shuanglin Road, Chengdu 610051, China}
\def\scKeyLab{Jinping Deep Underground Frontier Science and Dark Matter Key Laboratory of Sichuan Province, Liangshan 615000, China}

\def\pku{School of Physics, Peking University, Beijing 100871, China}
\def\CHEPpku{Center for High Energy Physics, Peking University, Beijing 100871, China}

\def\SDUdep{Research Center for Particle Science and Technology, Institute of Frontier and Interdisciplinary Science, Shandong University, Qingdao 266237, China}
\def\SDUlab{Key Laboratory of Particle Physics and Particle Irradiation of the Ministry of Education, Shandong University, Qingdao 266237, China}

\def\SYU{School of Physics, Sun Yat-Sen University, Guangzhou 510275, China}
\def\SYUSFI{Sino-French Institute of Nuclear Engineering and Technology, Sun Yat-Sen University, Zhuhai 519082, China}
\def\SYUzhuhai{School of Physics and Astronomy, Sun Yat-Sen University, Zhuhai 519082, China}
\def\SYUshenzhen{School of Science, Sun Yat-Sen University, Shenzhen 518107, China}

\def\NKU{School of Physics, Nankai University, Tianjin 300071, China}
\def\YTU{Department of Physics, Yantai University, Yantai 264005, China}
\def\FDU{Key Laboratory of Nuclear Physics and Ion-beam Application (MOE), Institute of Modern Physics, Fudan University, Shanghai 200433, China}
\def\CDUT{College of Nuclear Technology and Automation Engineering, Chengdu University of Technology, Chengdu 610059, China}

\affiliation{\tdli}
\author{Peiyuan Chen\orcidlink{0009-0008-8703-4495}}\affiliation{\sjtuphys}
\author{Wei Chen\orcidlink{0009-0009-5911-7135}}\affiliation{\sjtuphys}
\author{Xiaohua Chen}\affiliation{\tdli}
\author{Xun Chen\orcidlink{0000-0001-7961-7908}}\affiliation{\tdli}\affiliation{\SJTUSC}\affiliation{\scKeyLab}
\author{Yunhua Chen}\affiliation{\YaLongSD}\affiliation{\scKeyLab}
\author{Chen Cheng\orcidlink{0000-0003-0164-7538}}\affiliation{\BUAA}
\author{Xiangyi Cui}\affiliation{\tdli}
\author{Yuxin Cui}\affiliation{\BUAA}
\author{Manna Deng}\affiliation{\SYUSFI}
\author{Roni Dey\orcidlink{0000-0003-0513-9207}}\affiliation{\tdli}
% ...
\collaboration{PandaX Collaboration}
\noaffiliation